\documentstyle{mn}

\def\MSUN{\rm M_{\odot}}

\def\MSUNYR{\rm M_{\odot}\,yr^{-1}}
\def\MDOT{\dot{M}}

\newbox\grsign \setbox\grsign=\hbox{$>$} \newdimen\grdimen \grdimen=\ht\grsign
\newbox\simlessbox \newbox\simgreatbox
\setbox\simgreatbox=\hbox{\raise.5ex\hbox{$>$}\llap
     {\lower.5ex\hbox{$\sim$}}}\ht1=\grdimen\dp1=0pt
\setbox\simlessbox=\hbox{\raise.5ex\hbox{$<$}\llap
     {\lower.5ex\hbox{$\sim$}}}\ht2=\grdimen\dp2=0pt

\def\simless{\mathrel{\copy\simlessbox}}

\title{ Comparison of theoretical radiation-driven  winds 
from stars and discs.}

\author[D. Proga]{Daniel Proga \\ 
Imperial College of Science, Technology and Medicine, 
Blackett Laboratory, Prince Consort Road, London SW7 2BZ, UK \\
E-mail:  d.proga@ic.ac.uk } 

\begin{document}
\maketitle

\begin{abstract}
We compare models of  line-driven
winds from accretion discs and single spherical stars. 
We look at the problem of scaling mass-loss rates and velocities
of stellar and disc winds with model parameters. 
We find that stellar and disc winds driven by radiation,  
within the CAK framework, are very similar as far as mass-loss rates and 
velocities are concerned.
Thus we can use analytic results for stellar winds to rescale, 
in a first order approximation, numerical results for disc winds. 
We also show how the CAK stellar solutions change when  we take into account
effects of very low luminosities or line-driving force.

\end{abstract}


\section{Introduction} 

Radiation pressure has been long recognised as effective at powering mass-loss
from luminous stars. It has been shown that stars with luminosities
as low as 0.1 per cent of their Eddington limit can produce a powerful high 
velocity wind, when the transmission of radiation pressure to a flow via 
spectral line opacity is included. Studies of stellar winds driven by 
radiation pressure due to lines began more than two decades ago 
(Lucy \& Solomon 1970; Castor, Abbott \& Klein 1975, hereafter CAK). 
Later theoretical work by Friend \& Abbott (1986) 
and Pauldrach, Puls \& Kudritzki (1986; hereafter PPK) 
modified the CAK method  and reproduced well
empirically estimated time-averaged mass-loss rates and terminal velocities 
for main-sequence and evolved OB stars. After these successes of the modified 
CAK method, increasingly more attention has been paid to the instabilities 
inherent in the line-driving mechanism (e.g., Owocki, Castor \& Rybicki 1988, 
here after OCR; Puls, Owocki \& Fullerton 1993) and multi-dimensional aspects
of stellar winds (e.g., Bjorkman \& Cassinelli 1993; 
Owocki, Cranmer \& Blondin 1994).

A strong radiation field can also be produced by accretion discs, which
are believed to be important and common components in many astrophysical
objects, for example, cataclysmic variables (CVs), high
mass  young stellar objects (YSOs) and AGNs. Many of these objects
show evidence of high velocity winds coming from the disk itself
(e.g., Drew 1997; Mundt \& Ray 1994; Weymann et al. 1991).
Small wonder then, that radiation pressure has been proposed to power
mass-loss also from accretion discs.

Despite a deep and quantitative understanding of radiation-driven winds
in stars it has  not been straightforward to  demonstrate 
that radiation can produce strong and fast winds from discs. 
The basic difficulty is the intrinsically two-dimensional, axisymmetric
nature of disc winds. 
First studies of disc winds were focused on finding semi-analytic
time-independent solutions.
Such an approach required some simplifications and assumptions
which had an essential influence on the final outcome 
(e.g., Vitello \& Shlosman 1988; Murray et al 1995). 
Nevertheless these studies  illustrated in more detail  
the basic differences between driving 
a wind from a stellar photosphere and 
a disc photosphere.
For example, Vitello \& Shlosman showed that an increase of the
vertical  gravity component  with  height from the disc midplane  
prevents a disc from producing a wind unless the vertical
component of the radiation pressure also increases with height.

Numerical treatments of the disc wind problem have proven to be more
successful as they allow  solving of the multidimensional dynamical
equations from first principles.  Relatively early work by 
Icke~(1980; 1981) using two-dimensional, time-dependent numerical
simulations showed how  radiation pressure due to electrons  can produce
a disc wind. More recently Pereyra, Kallman \& Blondin (1997, see also 
Pereyra 1997) presented numerical calculations of the dimensional
structure of line-driven disc winds for CVs. However their spatial
resolution of  wind near a disc was too coarse to capture the wind structure
in that critical region.

Proga, Stone\& Drew (1998, hereafter PSD)  developed 
another numerical treatment of  disc winds. 
PSD have identified the inner and near disc as 
the important spatial domain and have 
therefore used a non-uniform (up to
200 $\times$ 200) grid to ensure that the subsonic acceleration zone
near the disc surface is well sampled.

PSD explored the impact upon the mass-loss rate and 
outflow geometry caused by varying the system luminosity and 
the radiation field geometry.
A striking outcome of PSD's study has been the finding that winds driven
from, and illuminated solely by, an accretion disc yield complex, unsteady
outflow. In this case, time-independent quantities can be determined only 
after averaging over several flow timescales.
On the other hand, if  winds are illuminated by radiation  dominated
by the central star, then the disc yields steady outflow.
PSD also found  that the mass-loss rate is  a strong function of 
the total luminosity,
while the outflow geometry is determined by the geometry of 
the radiation field. 
In particular, for high system luminosities, the disc mass-loss rate scales 
with the effective Eddington luminosity in a way  similar to stellar
mass loss. As the system luminosity decreases below a critical
value (about twice the Eddington limit), the mass-loss rate decreases 
quickly to zero.

PSD's calculations have been motivated by and designed for the case of winds
from CVs. Their findings are in  qualitative agreement with the kinematics
of CV outflows inferred from spectroscopic observations.  In 
a second paper (Drew, Proga \& Stone 1998; hereafter DPS) they considered 
both the star and  the disc as a source of mass. For model parameters
suitable for a typical high mass YSO, where the stellar luminosity
is $\sim 2$ orders of magnitude higher than the intrinsic disc luminosity, 
their  radiation-driven wind disc model  might explain the extreme
mass-loss signatures of these objects.
Most recently Oudmaijer et al. (1998) followed DPS and calculated
a  model for a B star surrounded by an optically thick disc. The 
resulting wind configuration could match with the wind structure
in B[e] stars. In particular, Oudmaijer et al.'s results
show that the kinematic properties of the wind might explain 
the observed spectral features in HD 87643.

In this paper we look at the problem of scaling disc mass-loss rates
and terminal velocities with model parameters. We will use results presented 
by PSD and DPS and use their approach to calculate some new results. 
Additionally we will review results for stellar winds to compare them  with 
numerical results for disc winds. For that we will calculate some new stellar 
wind models for a parameter space not explored before. These new stellar 
calculations are for relatively low luminosities, like those of 
white dwarfs and/or a very low increase of radiation pressure due to lines 
as in highly ionized\footnote{ 
Abbott \& Friend 1989 considered a related problem  of the effects of ionizing 
shocks in a stellar wind on the terminal velocity and mass-loss rate. Their 
approach was to assume that the radiation force is abruptly cut off at 
an adjustable distance from the star.} 
or low metallicity  winds.
We would like to point out that we treat our comparison between models of 
stellar and  disc winds quite formally as our goal is to identify  what 
and how model parameters control the mass-loss rate and velocity of 
disc winds.  The resulting scaling relations will allow a better 
understanding of line-driving in general and also give some tools 
to estimate disc  wind properties for many astrophysical applications 
before calculating detailed multidimensional time-dependent numerical models.

We describe our numerical calculations in Section 2, present and compare
our results in Section 3, and conclude with a discussion in Section 4.
The Appendix summarizes the analytic results for $\MDOT$ and 
the velocity law in the CAK framework.

\section{Numerical Methods}

\begin{table}
\footnotesize
\begin{center}
\caption{ 
Ranges of the model parameters which we have explored.}
\begin{tabular}{l l   } \\ 
\hline
  &    \\
Parameter & Range \\
  &    \\
\hline 
$c_s'$    & $1.53 \times 10^{-3} - 1.38 \times 10^{-2}$  \\
$\alpha$  & 0.4 -- 0.8 \\
$M_{max}$ &  1380 -- 4400   \\
$\Gamma_D$ & 0 -- $1.21 \times 10^{-2}$ \\
$\Gamma_\ast$ & 0 -- $1.21 \times 10^{-2}$ \\
\hline
\end{tabular}
\end{center}
\normalsize
\end{table}

To calculate wind models 
we adopt the 2.5-dimensional hydrodynamical numerical
method described by DPS (see also PSD). 
These models incorporate a non-rotating 
star,  and a geometrically-thin and optically-thick accretion disc. 
The star  and disc can both be  a source of radiation and mass.
PSD's formalism allows the stellar radiation to be included both
as a direct contributor to the radiation force and as an indirect component
via disc irradiation and re-emission. Each point on the disc and star
is assumed to emit isotropically. 
The  model takes into account stellar
gravity, gas pressure effects,  and rotational and radiation forces.
The gas in the wind is taken to be isothermal.
In the  dimensional form of the equations of motion for the wind
model studied by PSD there are six model parameters: 
the stellar Eddington number, 
$\Gamma_\ast=\frac{L_\ast \sigma_e}{4 \pi c G M_\ast}$, 
the disc Eddington number, $\Gamma_D=\frac{L_D \sigma_e}{4\pi c GM_\ast}$,  
($\Gamma_\ast=x \Gamma_D$ using the PSD $x$ parameter;
all other symbols have their conventional meaning), 
the normalised sound speed, $c'_s$, and 
three parameters describing the force multiplier, $\alpha$, $k$ 
and $\eta_{max}$ (see Appendix).
For parameters suitable for CVs (see PSD's table~1) 
the unit time, $\tau=\sqrt{{r_\ast^3}/(G M_\ast)}=2.88$s, the unit
velocity $v_0=\sqrt{G M_\ast/r_\ast}=3017~\rm km~s^{-1}$, 
the unit mass-loss rate, 
$\MDOT_0=8\pi c r_\ast/\sigma_e=2.62\times10^{-5}~\rm \MSUN~yr^{-1}$ and
$c_s'=c_s/v_0=4.6 \times 10^{-3}$.

Our calculations for stellar winds are as  described in DPS 
with some changes  outlined below. 
Because of the assumed spherical symmetry for stellar winds  we use
spherically symmetric initial and boundary conditions. 
We set all initial velocity components  to   zero,
except the radial component, $v_r$, for which we use the CAK
velocity (see equations A12 and A13). The density profile is given
by  hydrostatic equilibrium in the subsonic region and
the CAK density profile (i.e., using  the continuity equation, 
the CAK velocity law and mass-loss rate) in the supersonic region. 
The wind base density, $\rho_{lb}$, 
was chosen to sample the subsonic part of the flow
and to allow initial transients to disappear
on a short timescale. Typically, we use $\rho_{lb}=10^{-10}~{\rm g~cm^{-3}}$.
The radiation force is recalculated using
the method described in PSD for the case without a disc. 
Here we would like to mention
that in PSD, DPS and Oudmaijer et al. (1998) the radiation force due to lines
from a star was calculated using equation (A5), not PSD's equation (C2). 
Thus all previous calculations of those authors were with a finite 
disc correction.

To confirm and extend the PSD results, we present some additional disc 
models with $\alpha=0.4$, 0.6 and 0.8 calculated 
for various disc luminosities and with the stellar radiation switched off. 
The disc wind models are calculated  as described in PSD.

In numerical calculations with a finite line-force cut-off we used 
two approaches to saturate the force multiplier.
The first approach we considered is simply: 
\begin{equation}
M(t)= \left\{ \begin{array}{ll} 
k t^{-\alpha}
& {\rm for}~~\,~~ 
t > t_{min}\\
 & \\
 M_{max} 
& {\rm for} ~~\,~~
t \leq t_{min} \\
\end{array}
\right.
\end{equation}
where $t_{min}$ is chosen so $k t_{min}^{-\alpha}=M_{max}$.

Our second, more physical approach, is that $M(t)$ 
saturates gradually as  the wind material becomes more optically thin
(see also PSD). 
Specifically, we follow OCR who introduced  an exponential cutoff in the line
distribution which ensures the force multiplier saturates smoothly:
\begin{equation}
M(t)~=~k t^{-\alpha}~ 
\left[ \frac{(1+\tau_{max})^{(1-\alpha)}-1} {\tau_{max}^{(1-\alpha)}} \right]
\end{equation}
where $\tau_{max}=t\eta_{max}$ and $\eta_{max}$ is a parameter 
related to the most optically thick lines.
Equation (2) shows the following limiting behaviour:
\begin{eqnarray}
\lim_{\tau_{max} \rightarrow \infty}~M(t) & = & k t^{-\alpha} \\
\lim_{\tau_{max} \rightarrow 0}~M(t) & = & 
M_{max,e}= k (1-\alpha)\eta_{max}^\alpha.
\end{eqnarray}

For stellar winds, we calculated a few numerical test models to compare with
analytic results  summarized in the Appendix.
We found good agreement between  
the numerical and analytic results - typically better than a few per  cent.

\begin{figure*}
\begin{picture}(170,400)
\put(0,0){\includegraphics{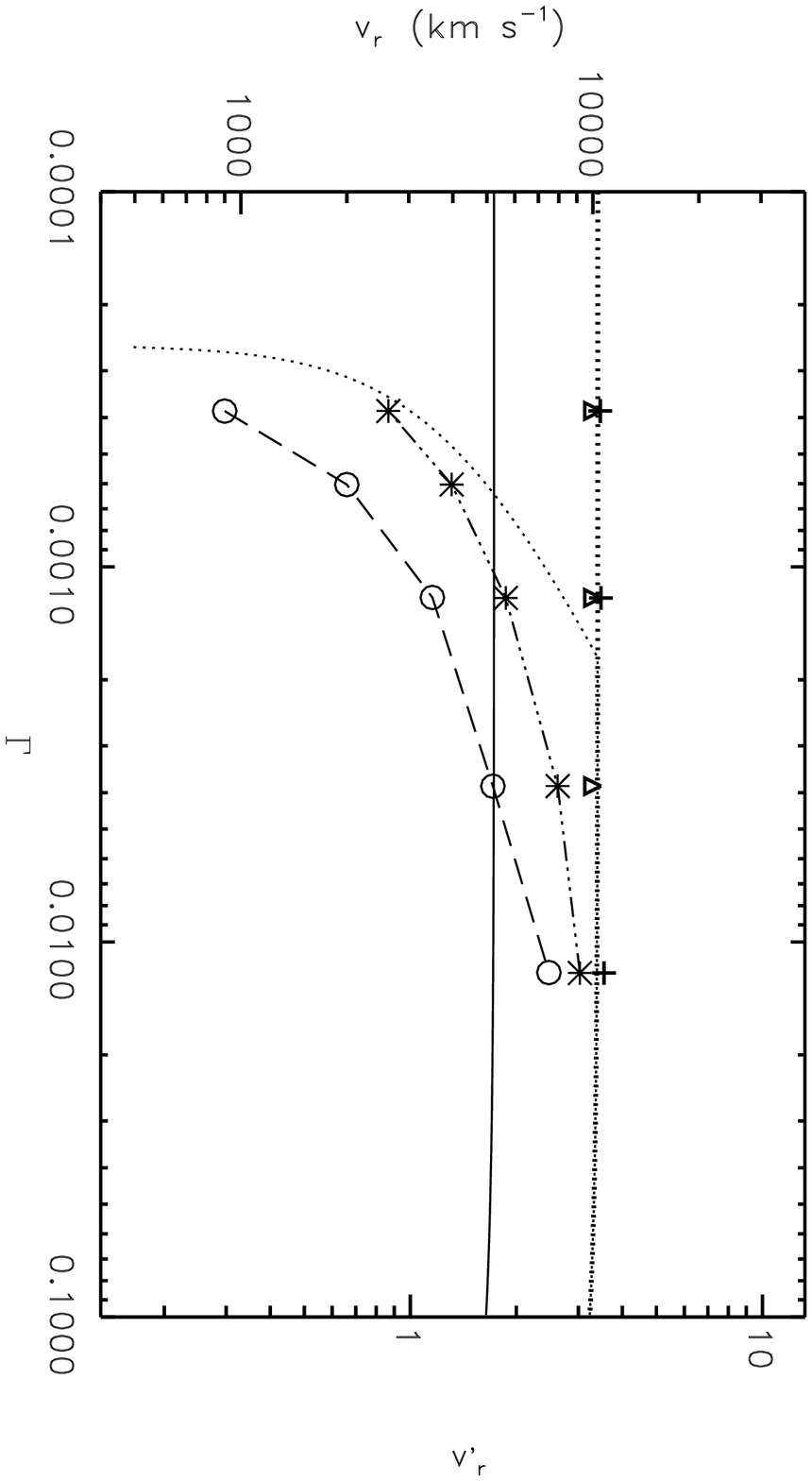}}
\put(0,0){\includegraphics{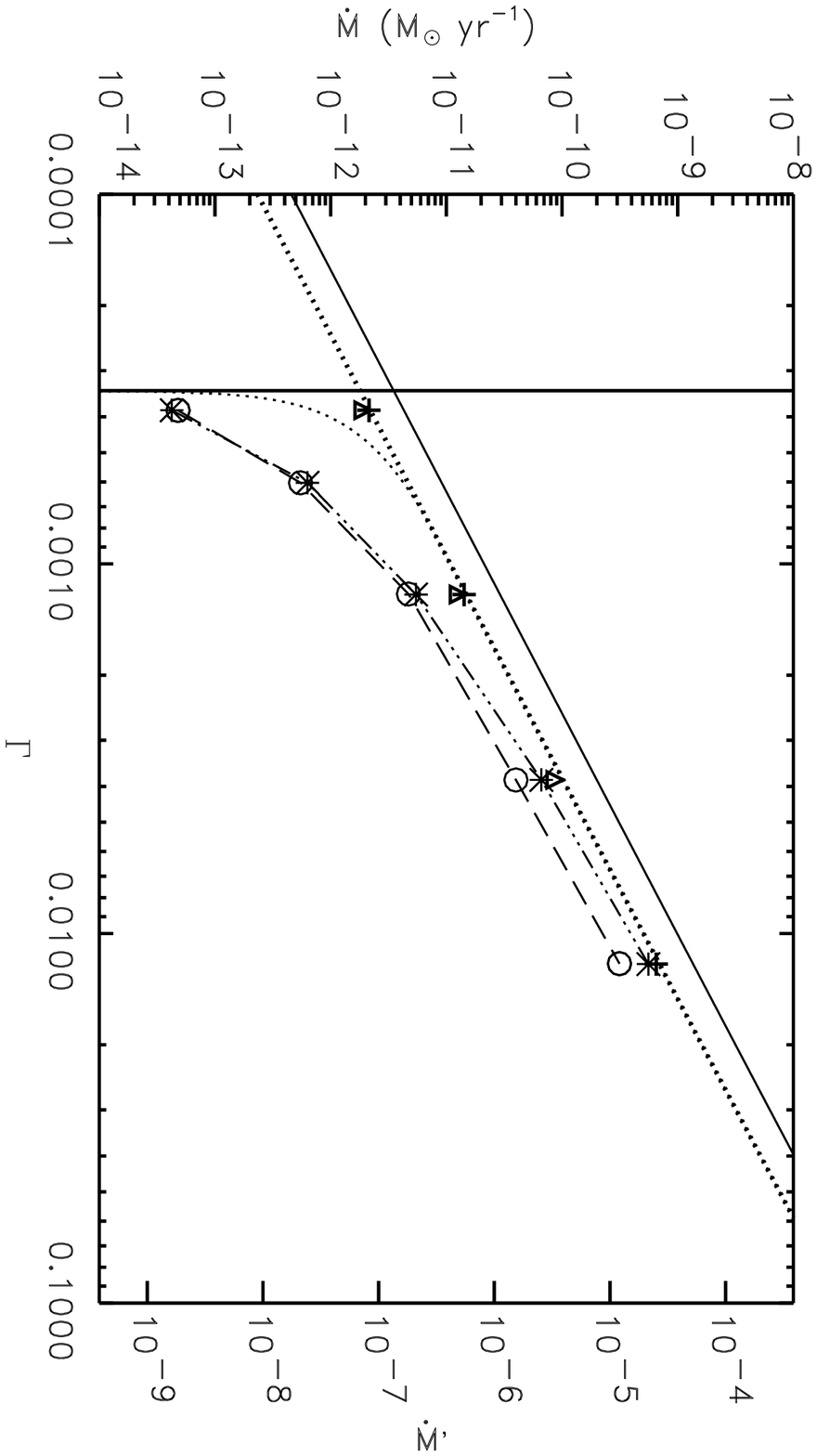}}
\end{picture}
\caption{ 
Model mass-loss rates (upper panel) and terminal/typical velocities 
(lower panel) as functions of the Eddington number.
The solid lines plot results for stellar models in the  CAK case 
(equations A13 and A14),
while the thin dotted lines plot results for  stellar models in  
the $CAK,FD\&M_{max}$ case (equations A25 and A26). 
The thick dotted lines plot results for stellar models in the CAK\&FD 
case (equations A18, A19, and A20) while the crosses  correspond 
to numerical results in this case.
The asterisks connected by the triple-dot dashed lines correspond to  
numerical results for stellar models in the $CAK,FD\&M_{max,e}$ case. 
The open triangles  represent disc results for models  in 
the $CAK\&FD$ case while the open circles, 
connected by the dashed lines, represent disc results 
in the $CAK,FD\&M_{max,e}$ case.
On upper panel the thick solid vertical line is 
for $\Gamma=1/(1+M_{max}f_{FD,c})$. The alternative ordinate on the
right hand side of the upper panel is the dimensionless wind mass-loss
rate parameter, $\dot{M}'=\MDOT/\MDOT_0$, and on the lower panel it 
is the dimensionless velocity parameter, $v'_r=v_r/v_0$.
}
\end{figure*}

\section{Results}

Table~1 lists the parameter ranges  explored in our numerical simulations.
Generally, $\Gamma_\ast$ and $\Gamma_D$ control all the other
parameters because they determine the wind photoionization structure
for a given chemical composition.
We thus  need to  explore only $\Gamma_\ast$ and $\Gamma_D$  provided we
calculate self-consistently the wind photoionization structure and subsequently
the radiation force. 
Unfortunately, this is 
not feasible  due 
to the wind's multidimensional nature and its potential time variability.
However  using models calculated for
stellar winds and some basic physical arguments we may consider 
coupling between $\alpha$, $\eta_{max}$, and 
$k$ without detailed photoionization calculations. 
Specifically, we are guided in this  matter by the results 
showed by Abbott (1982) and Stevens \& Kallman (1990) and
references therein.

\begin{table*}
\footnotesize
\begin{center}
\caption{ Summary of numerical results for stellar winds with 
$\alpha=0.6$, k=0.2 and $M_{max}=4400$.}
\begin{tabular}{ c c r  } \\ \hline 
                     &                   &           \\
 $\Gamma_\ast$       & $\MDOT_\ast$               &  $v_r(10 r_\ast)$\\ 
                     & (M$_{\odot}$ yr$^{-1}$)   & $(\rm km~s^{-1})$ \\ \hline  

                     &                 & \\

$CAK,FD\&M_{max}$  case         &                 &                \\
 $3.84 \times 10^{-4}$ & $ 9.0 \times 10^{-13}$   & 2770           \\
 $6.03 \times 10^{-4}$ & $ 4.1 \times 10^{-12}$   & 4550           \\
 $1.21 \times 10^{-3}$ & $ 1.4 \times 10^{-11}$   & 7640           \\
 $3.84 \times 10^{-3}$ & $ 9.6 \times 10^{-11}$   &11130           \\
 $1.21 \times 10^{-2}$ & $ 6.5 \times 10^{-10}$   &10750           \\ 
                       &                 & \\

$CAK,FD\&M_{max,e}$  case  &                 & \\
 $3.84 \times 10^{-4}$ & $ 4.2 \times 10^{-14}$  & 2630            \\
 $6.03 \times 10^{-4}$ & $ 6.3 \times 10^{-13}$  & 3970            \\
 $1.21 \times 10^{-3}$ & $ 5.5 \times 10^{-12}$  & 5660            \\
 $3.84 \times 10^{-3}$ & $ 6.6 \times 10^{-11}$  & 7940            \\
 $1.21 \times 10^{-2}$ & $ 5.6 \times 10^{-10}$  & 9180            \\ 

\hline
\end{tabular}

\end{center}
\normalsize
\end{table*}

Using results of previous radiative force calculations for single stars 
and arguments from atomic physics,  Gayley (1995) argues  
that the maximum force multiplier, $M_{max}$ is relatively constant and 
depends on metallicity only. Gayley found that $M_{max}\sim {\rm~a~few~} 10^3$.
Stevens \& Kallman (1990) calculated the radiative force for a stellar
wind ionized by an external X-ray source in massive X-ray binary systems.
Their predicted maximum force multipier is also a few $10^3$ 
and is nearly constant for low ionization parameter. 
For a very high photoionization parameter,
$M_{max}$ decreases to zero as all ions in a wind lose all electrons
and there is a reducing contribution to the radiation pressure from lines.
Stevens \& Kallman showed how the force multiplier decreases with the 
photoionization parameter. We will return to this point in Section 4.  

Table~2 summarizes our numerical results for stellar winds while
Table~3 summarizes PSD's results and ours for disc winds 
with  $\Gamma_\ast=0$.

\subsection{Dependence of wind properties on $\Gamma$ and $M_{max}$}

Previous  models of stellar winds driven by radiation force have
explored a wide range of model parameters  suitable for hot
luminous stars. In the Appendix we briefly describe 
the basics of the CAK method,
with the finite disc correction 
(hereafter CAK\&FD), and with explicit introduction
of $M_{max}$ (hereafter CAK,FD\&$M_{max}$). 
We also summarize the analytic results for $\MDOT_\ast$ 
and the velocity law.

In Fig.~1 we show (a) model mass-loss rates
and (b) the terminal/typical velocities  as  functions of the Eddington number.
All models on Fig.~1 are  for  $\alpha=0.6$, $k=0.2$ and $M_{max}=4400$. 
We plot analytic results for the stellar mass-loss rate, 
$\MDOT_\ast$, and the terminal velocity, $v_\infty$ 
in four cases:  $CAK$,   $CAK\&FD$, $CAK,FD\&M_{max}$, and
$CAK,FD\&M_{max,e}$. 
The last case, $CAK,FD\&M_{max,e}$ is where the finite disc factor and 
OCR's exponential cutoff in the line distribution are included.

For high $\Gamma_\ast$, the results shown on Fig.~1(a) confirm the well known CAK 
relation $\MDOT_\ast \propto \Gamma_\ast^{1/\alpha}$
(the thin solid and thick dotted lines).
Additionally our results confirm that  the inclusion of the finite disc 
correction reduces $\MDOT_\ast$ by a factor of 
$\sim 2 \approx (1+\alpha)^{1/\alpha}$, for $\alpha \sim 0.6$ 
(Friend \& Abbott 1986 and PPK). This figure also reveals that  
the CAK scaling does not apply for very small $\Gamma_\ast$ if 
we take into account the fact that the force multiplier can not increase to 
arbitrarily high value (the thin dotted line and the asterisks).
Our analysis presented in Appendix shows that the CAK scaling does 
not apply for 
\begin{equation}
\Gamma_\ast \leq 1/(1+(1-\alpha)f_{FD,c}M_{max}). 
\end{equation}
The mass loss vanishes at 
\begin{equation}
\Gamma_\ast=1/(1+M_{max}f_{FD,c}).
\end{equation} 
Previous studies by other authors of line driven stellar winds 
were for luminous stars for which the presence of $M_{max}$ was not important. 
However for $\Gamma_\ast\simless0.01$ in the $CAK,FD\&M_{max}$ case, 
$\MDOT_\ast$ is lower and the $\MDOT_\ast$ vs. $\Gamma_\ast$ relation 
is steeper than in the $CAK\&FD$ case. 
For $\Gamma_\ast>0.01$, $\MDOT_\ast$ in the $CAK,FD\&M_{max}$ case converges
with $\MDOT_\ast$ in the $CAK\&FD$ case. 
On the other hand, $\MDOT_\ast$ in the $CAK,FD\&M_{max,e}$ case is steeper
than in the $CAK,FD\&M_{max}$ case.  Additionaly, 
$\MDOT_\ast$ in the $CAK,FD\&M_{max,e}$ case converges with 
$\MDOT_\ast$ in the $CAK\&FD$ case for higher $\Gamma_\ast$ than with
$\MDOT_\ast$ in the $CAK,FD\&M_{max}$ case.

\begin{table*}
\footnotesize
\begin{center}
\caption{ Summary of parameter survey for disc winds with $k=0.2$ and 
$M_{max}=4400$.}
\begin{tabular}{c c c r c  } \\ \hline 
          &             &             &                   &          \\
 $\alpha$ & $\Gamma_D$  & $\MDOT_D$   &  $v_r(10 r_\ast)$ & run number \\ 
          &             &  (M$_{\odot}$ yr$^{-1}$) & $(\rm km~s^{-1})$ & in PSD \\ \hline  

          &             &             &                       &     \\
 $CAK,FD$ case          &             &             &                       &     \\
 0.6      &$3.84 \times 10^{-4}$  &$ 1.9\times10^{-12}$   & 10000  &   \\
 0.6      &$1.21 \times 10^{-3}$  &$ 1.3\times10^{-11}$   & 10000  &   \\
 0.6      &$3.84 \times 10^{-3}$  &$ 8.7\times10^{-11}$   & 10000  &   \\

          &             &             &                       &     \\
 $CAK,FD\&M_{max,e}$ case          &             &             &                       &     \\
 0.4      &$3.84 \times 10^{-4}$  &$ 2.0\times10^{-17}$   &  700 &  \\
 0.4      &$6.03 \times 10^{-4}$  &$ 6.3\times10^{-16}$   & 1400 &  \\
 0.4      &$1.21 \times 10^{-3}$  &$ 6.3\times10^{-15}$   & 3000 & 15 \\
 0.4      &$3.84 \times 10^{-3}$  &$ 1.6\times10^{-13}$   & 6000 & 16 \\
 0.4      &$1.21 \times 10^{-2}$  &$ 2.4\times10^{-12}$   &10000 &  \\ 

          &             &             &                       &     \\
 0.6      &$3.84 \times 10^{-4}$  &$ 4.8\times10^{-14}$   &  900 & 2 \\
 0.6      &$6.03 \times 10^{-4}$  &$ 5.5\times10^{-13}$   & 2000 &   \\
 0.6      &$1.21 \times 10^{-3}$  &$ 4.7\times10^{-12}$   & 3500 & 3 \\
 0.6      &$3.84 \times 10^{-3}$  &$ 4.0\times10^{-11}$   & 5200$^{(a)}$ & 4 \\
 0.6      &$1.21 \times 10^{-2}$  &$ 3.1\times10^{-10}$   & 7500 & 5 \\ 

          &             &             &                       &      \\
 0.8      &$3.84 \times 10^{-4}$  &$ 1.9\times10^{-13}$   &  500 &   \\
 0.8      &$6.03 \times 10^{-4}$  &$ 8.8\times10^{-12}$   & 1200 &   \\
 0.8      &$1.21 \times 10^{-3}$  &$ 6.2\times10^{-11}$   & 2500 & 17 \\
 0.8      &$3.84 \times 10^{-3}$  &$ 6.3\times10^{-10}$   & 7500 & 18 \\
 0.8      &$1.21 \times 10^{-2}$  &$ 4.0\times10^{-9} $   & 14000 & 19 \\ 

\hline
\end{tabular}

\end{center}
\normalsize
a) We re-estimated a typical velocity for this model.
\end{table*}

Fig.~1(a) also shows  numerical results for the disc mass-loss rate, $\MDOT_D$,
from PSD (with a new point for $\Gamma_D =6.03\times10^{-4}$, see Table~3), 
and for disc wind models with infinite $M_{max}$, the open circles and 
triangles, respectively.
PSD results are  analogous to stellar results in the $CAK,FD\&M_{max,e}$ case,
while disc results with  infinite $M_{max}$ are analogous to
stellar results for the $CAK,FD$ case.
Comparison between these analogue models  demonstrates
that the stellar mass-loss rate  and the disc mass-loss rate depend 
on luminosity in a similar way for all luminosities. 
The most convincing case is for disc wind models with infinite $M_{max}$ 
where disc mass-loss rate predictions almost match the  results for
the stellar $CAK\&FD$  case. 
In the $CAK,FD\&M_{max,e}$ case, lower $\MDOT_D$ approaches 
$\MDOT_\ast$ in the stellar
$CAK,FD\&M_{max}$ and $CAK\&FD$ cases as  the Eddington number increases.

Fig.~1(b) illustrates the behavior of  $v_\infty$
with the Eddington number. 
The CAK theory predicts that $v_\infty$ stays almost
constant with $\Gamma_\ast$ for $\Gamma_\ast \ll 1$
(see equation A13).  Friend \& Abbott (1986) showed that 
the finite disc factor introduces
a dependence of $v_\infty$ on $\Gamma_\ast$ but  it
is a weak dependence. Our calculations confirm these results
and also confirm that $v_\infty$ is higher for the $CAK\&FD$  case 
than for the $CAK$ case by a factor of $\sim 2$ (Friend \& Abbott 1986; PPK).
However  if we take into account $M_{max}$, $v_\infty$ decreases 
strongly with decreasing $\Gamma_\ast$ (see also equation A22).

Comparing results shown in Fig.~1 , 
we find that for the $CAK,FD\&M_{max}$ case
$v_\infty$ remains sensitive to  $M_{max}$ for higher $\Gamma_\ast$ 
than $\MDOT_\ast$.
Quantitatively   $v_\infty$ in the $CAK,FD\&M_{max}$ case
starts to deviate from $v_\infty$ in the $CAK\&FD$ case
at 
\begin{displaymath}
\Gamma_\ast=1/(1+2(1-\alpha)f_{FD,c}M_{max}
\end{displaymath}
while $\MDOT_\ast$ starts to deviate at
\begin{displaymath}
\Gamma_\ast=1/(1+(1-\alpha)f_{FD,c}M_{max}, 
\end{displaymath} 
higher than the former by a factor of $\approx 2$. 
This is  unsurprising, given that
the terminal velocity of stellar winds is a  global quantity 
determined in the whole  acceleration zone whereas $\MDOT$ is 
a local  quantity determined at the critical point and conserved
in steady state models (see also Abbott \& Friend 1989).

The characteristic velocity of disk winds does not depend on 
the Eddington number
for infinite $M_{max}$. However for $M_{max}=4400$, 
the characteristic velocity is a strong function of $\Gamma_D$.
Thus the disk wind velocity and the stellar wind velocity
are sensitive to the Eddington number in  similar ways.
The main difference is that
disc winds react to finite $M_{max}$ for higher Eddington number 
than stellar winds. This difference  indicates that $M(t_c)$ is higher 
for discs than for stars.

So far we showed that a finite $M_{max}$  reduces $\MDOT_\ast$ and $v_\infty$
significantly unless $M(t_c)< M_{max}$.  
In the regime where $M(t_c)<M_{max}$ the actual value of $M_{max}$
does not matter and $\MDOT_\ast$ and $v_\infty$
do not depend on $M_{max}$ nor  $\eta_{max}$.
As CAK found the main stellar properties are determined at the wind critical
point. For example, equation (A18) shows 
how the mass-loss rate  depends on the force multiplier at the critical point. 
We can also express
wind terminal velocity  as  a function of  $M(t_c)$:
\begin{equation}
v^2_\infty \propto \frac{(1-\Gamma_\ast)}{(1-\alpha)}=M(t_c) \Gamma_\ast
\end{equation}
(see the Appendix).
Thus luminous stars do not use all their available effective radiative force 
to drive winds. 
However for low luminosity stars, the importance of $M_{max}$ and the effective
luminosity is essential as $M_{max}$ limits $M(t_c)$. 
For example,   equation (A22) shows that 
in a first order approximation $v^2_\infty \propto M_{max}\Gamma_\ast$
(see equation 7).
Physically, it means that if the force multiplier is saturated
the radiation pressure accelerates the wind maximally
which is proportional to  $M_{max}\Gamma_\ast$, 
in the single-scattering limit. Additionally, 
as we showed  in Fig.~1  for the stellar case and PSD showed 
for the disc case, radiation pressure cannot drive a wind 
if $M_{max} \Gamma  < 1/f_{FD,c}\simless 1.6 $, 
that is if the effective luminosity is less than the Eddington luminosity.

To  confirm further that disc winds behave in the same way as stellar winds
we calculated a test model to compare
with PSD's fiducial steady model for $\Gamma_\ast=\Gamma_D=1.21 \times
10^{-3}$ (their model 8). 
For the test model all parameters are as for
the fiducial model except for $\Gamma_D=\Gamma_\ast=3.84 \times 10^{-3}$,  
increased  by a factor
of $10/\pi$, and $M_{max}=1380$  decreased by the same factor.
With these parameters our test model has  $M_{max}\Gamma$ 
the same as the fiducial model and the Eddington numbers as PSD's model 9.

The results of our test model confirm  expectations:
the test model has $\MDOT$ as PSD's model 9 with the same $\Gamma$
and $v_r(r_o)$ as PSD's 
model 8\footnote{Note a typo in PSD's  table 2: for model 8, $\MDOT_D$ 
should be $2.1 \times 10^{-15}~\MSUNYR$ instead of 
$1.2 \times 10^{-15}~\MSUNYR$.}
with the same $\Gamma M_{max}$. The test model
also settles into a steady state.
Then  the disk mass-loss rate, as the stellar mass-loss rate, 
depends on the Eddington
number while velocity depends on the effective luminosity 
at the critical point.


\subsection{Dependence of wind properties on~$\alpha$~and~$k$}

In the stellar case for $M(t_c)<M_{max}$, the analytic solutions provide 
scaling of the wind properties on all model parameters.
For example, the $\alpha$ parameter determines a slope of
the power-law dependence of $\MDOT_\ast$ on luminosity and on $k$ (i.e., 
$\MDOT_\ast \propto (k \Gamma_\ast)^{1/\alpha}$).

\begin{figure*}
\begin{picture}(170,400)
\put(0,0){\includegraphics{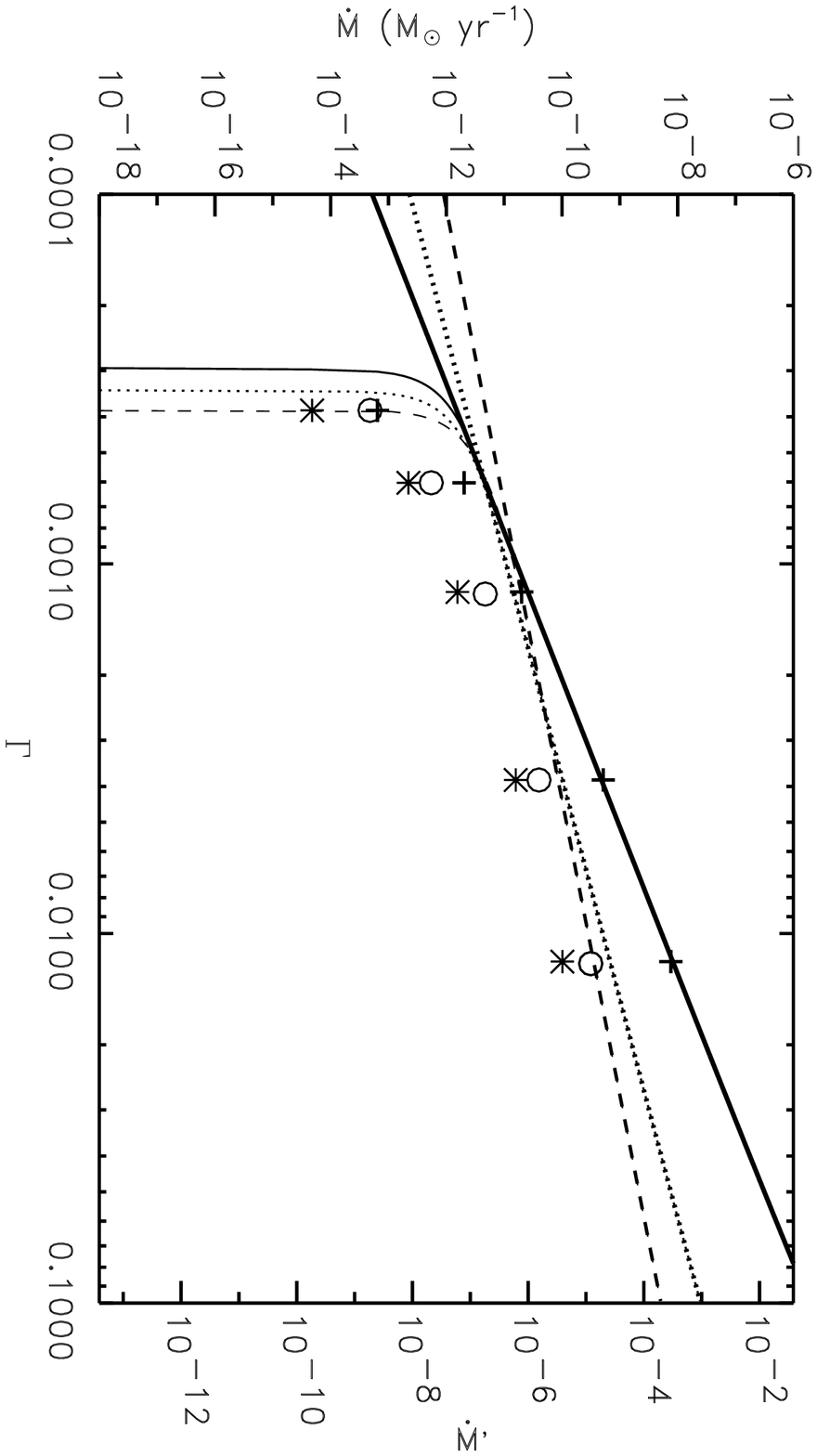}}
\put(0,0){\includegraphics{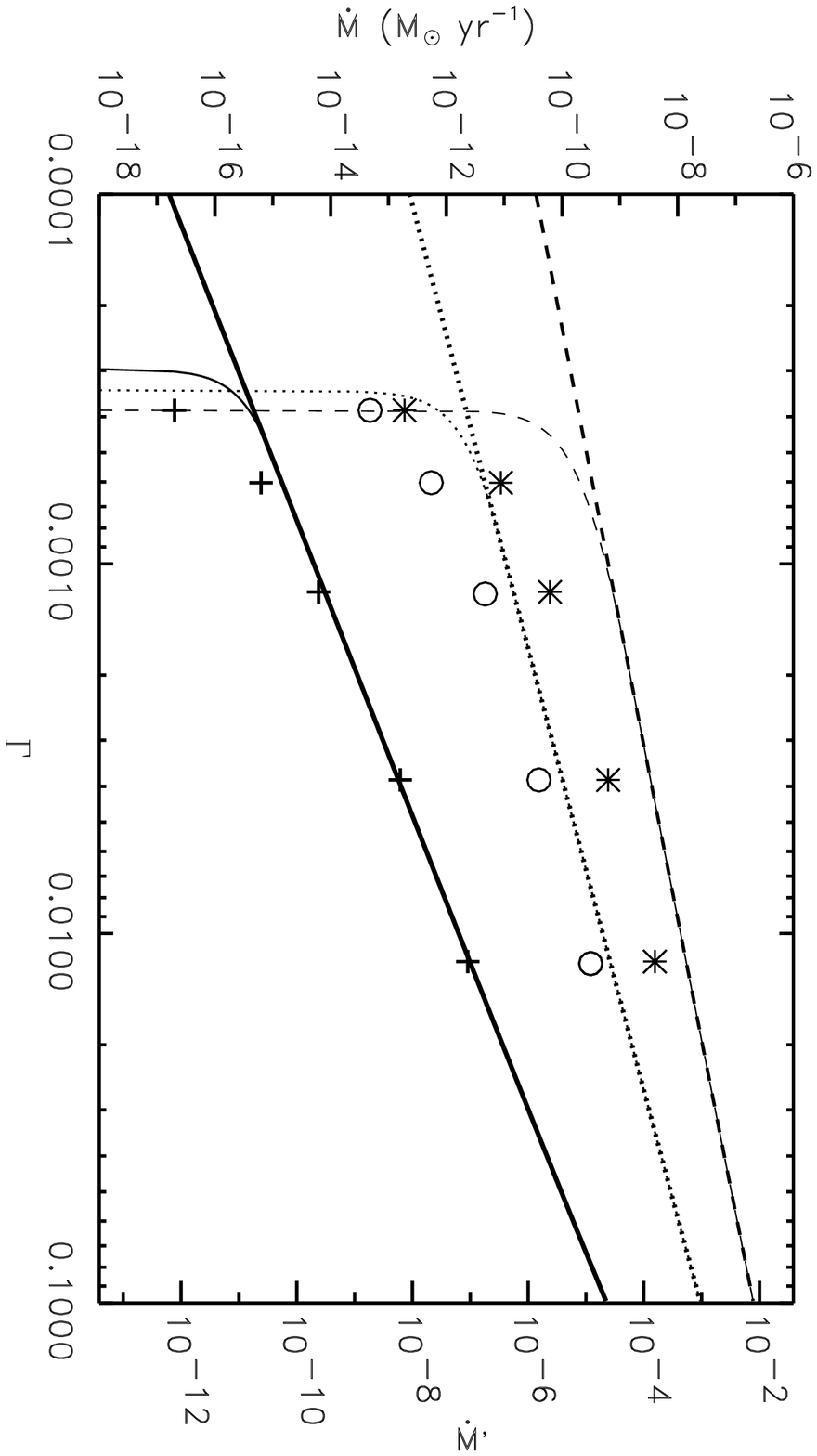}}
\end{picture}
\caption{ Model mass-loss rates as function of the Eddington number
for $k=0.2$ (upper panel) and k rescaled with $\alpha$ 
as described in Section 3.2 (lower panel). 
The thick lines represents results in the stellar
$CAK\&FD$ case for $\alpha=0.4, 0.6$ and 0.8; the solid, dotted, and
dashed lines, respectively. The thin lines represents
results in the stellar $CAK,FD\&M_{max}$ case, the line types
are as in the $CAK\&FD$ case. 
The open circles represent results for the disc $CAK,FD\&M_{max,e}$ case 
and $\alpha=0.6$, while the crosses and asterisks represent results for 
$\alpha=0.4$ 
and $\alpha=0.8$, respectively. The alternative ordinate on the right
hand side of the two panels is as in Fig.~1(a).
}
\end{figure*}

For $M(t_c)>M_{max}$ in the $CAK,FD\&M_{max}$ case,
$\alpha$ and $k$ determine the slope of a {\it linear} 
dependence of $\MDOT_\ast$ on luminosity:
$\MDOT_\ast \propto \Gamma k^{1/\alpha}$. Additionally, $\alpha$
enters the expression for $M(t_c)$ and then determines the lowest value
of $\Gamma$ for which $M(t_c)<M_{max}$ and the power-law applies 
(see equation 5). Moreover, through $f_{FD,c}$ the $\alpha$
parameter defines the lower limit of $\Gamma$ for which
a wind solution exists (see equation 6). 

Fig.~2 illustrates how the mass-loss rate  
depends on the  Eddington number for various $\alpha$ with (a) $k=0.2$
and (b) $k$ rescaled with $\alpha$ (see below).
Fig.~2 also compares our predictions for $\MDOT_\ast$ and $\MDOT_D$.
We do not consider the $CAK,FD,\&M_{max}$ case with $\alpha=0.4$ 
and 0.8 because our calculations for $\alpha=0.6$ are sufficient 
to show the basic behavior in this case.

In the disc case, we used PSD's results complemented by our results.
PSD calculated  a few disc wind models with $\alpha=0.4$, 0.6 and 0.8
for various disc and stellar luminosities in the $CAK,FD\&M_{max,e}$
case. They found that there is no change in the power-law dependence of 
$\MDOT_D$ on luminosity between $\alpha=0.6$ and $\alpha=0.8$, 
in contrast to scaling found for the stellar wind solutions.

The analysis in previous subsection showed that $\MDOT_D$ scales 
with the Eddington number  for $\alpha=0.6$ almost as $\MDOT_\ast$.
However the analysis also showed
that  the CAK scaling does not apply in both the disc and stellar cases
if  $\Gamma$ is very low and a finite $M_{max}$ is used. 
Using  these conclusions about similarities between disc and stellar winds,  
a simple interpretation of the PSD results for $\alpha=0.8$ is
that for these disc models $M(t_c)< M_{max}$ and the power-law 
does not apply.
As we mentioned above,
$M(t_c)$ increases with $\alpha$ (equation A17)
and subsequently the lowest $\Gamma$ for which the power-law  
applies,  increases to unity with $\alpha$,
(equation 1).  Then for a fairly high $\alpha=0.8$, 
even the highest-considered $\Gamma$ in Fig.~2 might still be in the region
where the power-law does not apply in the disc $CAK,FD\&M_{max,e}$ case. 
If so, no change in the power-law found by PSD and confirmed here can be seen 
as  cancellation of the two opposite trends
1) a decrease of the slope caused by higher $\alpha$ and 
2) an increase of the slope caused
by a finite $M_{max}$ (see Fig.~1).

This interpretation is confirmed by the  results for disc winds with 
$\alpha=0.4$.
Fig.~2 shows that for $\alpha=0.4$, the slope is as predicted by 
the CAK\&FD scaling and the power-law applies for  a Eddington
number range wider 
than for $\alpha=0.6$ and $\alpha=0.8$.

In their preliminary parameter survey, 
PSD  calculated  disc wind models for 
different $\alpha$ and $\Gamma_D$  while holding $M_{max}$ and $k$ fixed. 
In their formalism, they did that by means of adjusting a value of 
$\eta_{max}$ 
according to the equation
\begin{equation}
\eta_{max}=\left(\frac{M_{max}}{k(1-\alpha)}\right)^{1/\alpha}.
\end{equation}
As we mentioned in Section 3.1 Gayley (1995) showed that $M_{max}$ 
can be interpreted as 
a pseudo-conserved quantity that is proportional to metallicity and is 
estimable a priori. However $k$ depends on metallicity, sound speed and 
the other force multiplier parameters (e.g., CAK, Gayley 1995). Assuming 
that $M_{max}$ is constant and using equation (4) we can express $k$ as
\begin{equation}
k= \frac{1}{1-\alpha}\eta_{max}^{-\alpha} M_{max}.
\end{equation}
Additionally adopting Gayley's working assumption that  
we truncate the line distribution at the single strongest line 
(i.e., $M_{max}=\frac{v_{th}}{c}\eta_{max}$  using our notation)
we can rewrite equation (9) as
\begin{equation}
k= \frac{1}{1-\alpha}\left(\frac{v_{th}}{c}\right)^{\alpha} M_{max}^{1-\alpha}.
\end{equation}
The last equation shows explicitly how the formal parameter
$k$ depends on physical properties of wind.

In this context we could ask how PSD's and our results for fixed $k$ and 
$M_{max}$ can be rescaled to accommodate the coupling of $k$ with other model
parameters.

Using the result for stellar winds: $\MDOT\propto k^{1/\alpha}$, we can 
rescale the mass-loss rates
for a new k parameter, $k_n$ if we  multiply $\MDOT$ by the correction
factor:
\begin{equation}
f_k=\left(\frac{k_n}{k}\right)^{1/\alpha}.
\end{equation}   

The detailed calculations of the force multiplier and stellar winds 
(e.g., CAK; Abbott 1982; Gayley 1995) showed that the mass-loss rate 
increases with decreasing $\alpha$. This result can be understood
if we consider the coupling between the parameters of the force multiplier.
As equation (9) shows, $k$ increases with decreasing
$\alpha$ causing an overall increase of $M(t)$ 
and subsequently an increase of $\MDOT$.

PSD's results  suggest that $\MDOT_\ast$ increases with $\alpha$. 
This result is the consequence of changing $\alpha$ but 
formally keeping $k$ constant.
We found a similar behaviour of $\MDOT_\ast$ (see Fig. 2a) when we applied
PSD's approach.
However our previous analysis and numerical tests show that we can rescale 
$\MDOT_D$ if other $k$ is chosen. For example, if we take PSD models for
$\alpha=0.4$ and 0.8 and want to rescale them, 
keeping $M_{max}$ and $\eta_{max}$ as for  $\alpha=0.6$ 
(i.e., $M_{max}=4400$ and $\eta_{max}=10^{7.9}$) 
we need  first calculate $k_n$ using equation (9)
and then use the correction factor $f_k$. In the example considered here
the correction factor is:
\begin{equation}
f_k=\left(\frac{4400}{0.2(1-\alpha)}\right)^{1/\alpha} 10^{-7.9}.
\end{equation}   
Fig.~2b shows results from Fig.~2a, where the mass-loss rates 
for $\alpha=0.4$ and 0.8
are rescaled as described. The results for $\alpha=0.6$ 
are unchanged. Fig.~2a clearly shows that after rescaling $\MDOT_D$  
and $\MDOT_\ast$ 
increase with decreasing $\alpha$ as found in previous  stellar wind models.

Contrary to the mass-loss rate, the terminal velocity
does not depend on k (e.g., equations 
A13, A22, and A25). Our numerical tests confirm this for stellar and
disk winds.

\subsection{Dependence of wind properties on sound speed }

Analytic steady state solutions of stellar winds depend very weakly on 
the sound speed (e.g., CAK; PPK). 
Numerical simulations of stellar winds
showed that a higher $c_s$ shortens the computational time needed for the
initial conditions to settle into a steady state (e.g., OCR). However
the final steady state solution does not depend on $c_s$.
Our numerical tests confirm these results for stellar winds.
We thus do not expect changes of our disc wind results with $c_s$ either.
Nevertheless we calculated a few models for various $c_s$.

We have calculated models with  $c'_s$  higher and lower 
than $c_s'=4.6 \times 10^{-3}$ by a factor of 3 
and compare them with  PSD's fiducial complex and steady
models, PSD's model 2 and 8 respectively. We found that
$\MDOT_D$ and velocity of the disc winds do not depend on $c_s'$.
However, the time behaviour of the winds does depend.
For example, the PSD fiducial steady state model 
recalculated with $c_s'= 1.53 \times 10^{-3}$ is somewhat  time dependent: 
density fluctuations originating in the wind base spread in the form 
of  streams sweeping outward. We did not find infalls of dense clumps 
of gas on the disc.
On the other hand, the model as the PSD fiducial complex model but with 
$c_s' = 1.38 \times 10^{-2}$ settles into much smoother and 
steadier state than PSD's model 2.
Generally, we found that wind time dependence weakens with increasing
$c_s'$ because 
the gas pressure effects get stronger with increasing $c_s'$. 
Subsequently higher gas pressure  smooths the flow more effectively
and in a larger region above the disc mid-plane
as the size of the subcritical part of the flow increases with $c'_s$.

\section{Discussion}

Our comparison between disc and stellar winds has shown that despite
differences in geometry,
these two winds are similar in terms of $\MDOT$ and $v_\infty$.
In particular, we showed that $\MDOT_\ast$ is an 
upper limit on $\MDOT_D$ for the same model parameters,
typically $\MDOT_\ast/\MDOT_D\simless 2$.

Previous CV  wind simulations showed that the predicted
$\MDOT_D$ is lower than the observed $\MDOT_D$ by a factor
of a few, at  best (e.g., Drew 1997, PSD and references therein).
If we use $\MDOT_\ast$  for the Eddington number and other
model parameters as for CV's accretion discs we still may not
reproduce observed disc mass-loss rates. Thus we might conclude that
1) present disc wind models  still need some important improvement
or 2) $\MDOT_D$ inferred from observations are overestimated, or both.
At the moment redetermination of $\MDOT_D$ from observations taking into
account wind structure predicted by PSD is probably better
than moving directly to exploring
different ways of enhancing theoretical $\MDOT_D$ (e.g., including
magnetic field effects). Such an approach will provide better constraints
on our disk wind model.

Our tests with different sound speeds showed that 
thermodynamic effects might be important in determining 
time behavior of the disc winds.
However we do not believe that a sound speed
higher by a factor of 3 compared to  the one used  PSD for CVs 
is possible because
this could correspond to the wind temperature higher by a factor of 9!
Thus we do not expect great changes in models for CV disc winds after the
inclusion of temperature calculations.

Thermodynamics and changes in wind photoionization might be important
in very luminous systems such as AGNs and X-ray binaries.
In studies of the former there is the long standing problem of wind 
over-ionization by the central radiation, especially in the inner part of 
an outflow.
Such over-ionization  will manifest in a low $M_{max}$, for instance. 
However it is arguable
how damaging for radiation driving that can be. Using  fiducial values
for  AGN's: $\Gamma=0.4$ and  $\alpha=0.6$ (e.g., Murray et al. 1995; 
Arav 1996) and equation (A19), $M(t_c)=3.75$. Our analysis showed that line driven winds
are strong and fast when $M_{max}~>~M(t_c)$. Thus for AGNs to produce
strong and fast line-driven winds not too many absorption lines are needed.
Stevens \& Kallman (1990) showed that $M_{max} < 4$  for relatively
high photoionization   parameters (see also 
Arav, Li\& Begelman 1994; Murray~et~al. 1995).
This simple argument  supports the idea that winds in AGNs are line-driven.

To conclude we would like to mention that previous attempts to model disc winds
using 1D approximations, have found some vindication in 
our fully 2D time-dependent simulations.
This is not surprising, because of the energy argument: 
in both the stellar and disc cases most of the radiation energy is
released near the central object.
Thus  the fully 2D calculations are  crucial to determine
the density and velocity as function of the position 
and time but less necessary to estimate the mass-loss rate and
typical velocity for disc winds.

{\bf Acknowledgments:} We would like to thank Janet Drew and James Stone 
for helpful discussions. 
Scott Kenyon is thanked for his comments and 
careful reading of the manuscript.
This research has been supported by a research
grant from PPARC.  Computations
were performed at the Pittsburgh Supercomputing Center and 
Imperial College Parallel Computing Centre.

\onecolumn
\appendix

\section{Summary of analytic results  
for  the CAK model with the finite disc correction
and a finite maximum force multiplier.}

The radial radiation force (per unit mass) due to spectral 
lines at every location in the flow can be expressed as 
\begin{equation}
    F^{rad,l} = \int \left(\frac{\sigma_e d{\cal F}}{c}\right) M(t),
\end{equation}
where the term in brackets  is the radiation force due to electron 
scattering from a stellar surface element and ${d\cal F}_\ast$ is 
the frequency integrated flux from that element. The integration is over 
the visible radiant surface. The quantity $M$, called the force multiplier, 
represents the increase in radiation force over the pure electron scattering 
case when lines are included.   CAK found a fit to their numerical results 
for the force multiplier, such that
\begin{equation}
M(t)~=~k t^{-\alpha},
\end{equation} where $k$ and $\alpha$ are parameters of the fit.
In the  Sobolev approximation, $M(t)$ is
a function of the optical depth parameter
\begin{equation}
t~=~ \sigma_e \rho v_{th} \left|\frac{dv_l}{dl}\right|^{-1},
\end{equation}
where $\rho$ is the density, $v_{th}$ is the thermal velocity, and
$\frac{dv_l}{dl}$ is the velocity gradient along the 
direction linking a point in the wind to the radiant surface element.
For a spherically-expanding wind,
\begin{equation}
\frac{dv_l}{dl}= \frac{dv_r}{dr} \mu^2 +\frac{v_r}{r} (1-\mu^2),
\end{equation}
where $v_r$ is the radial wind velocity and $\mu$ is the direction cosine.

Making an assumption $\frac{dv_r}{dr} \gg \frac{v_r}{r} $, certainly valid
in a lower part of the wind acceleration zone, $\frac{dv_l}{dl}$ is
\begin{equation}
\frac{dv_l}{dl}= \frac{dv_r}{dr} \mu^2.
\end{equation}
Then using equations (A3) and (A4), the equation for $F^{rad,l}$ is
\begin{equation}
    F^{rad,l} = \left(\frac{\sigma_e {\cal F_\ast}}{c}\right) f_{FD} M_{CAK}(t),
\end{equation}
where $\cal F_\ast$ is the frequency integrated flux from the star,
$M_{CAK}(t)~=~ k \left(\sigma_e \rho v_{th} \left|\frac{dv_r}{dr}\right|^{-1}
\right)^{-\alpha}$,
is the force multiplier in the form introduced by CAK,
and $f_{FD}$ is the so-called finite disc correction factor
which allows a generalization of the CAK formalism beyond the point-star
case. The assumption made above simplifies the finite disc factor
to a function of radius only:
\begin{equation}
    f_{FD}(r) = \frac{1-(1-r_\ast^2/r^2)^{\alpha+1}}{(1+\alpha)r_\ast^2/r^2}.
\end{equation}

Using the Eddington number, 
$\Gamma_\ast=\frac{L_\ast\sigma_e}{4\pi c GM_\ast}$, where
all symbols have their conventional meaning, we can rewrite
equation (A6) as
\begin{equation}
    F^{rad,l} = \frac{G M \Gamma_\ast}{r^2} f_{FD}(r) k t^{-\alpha}.
\end{equation}

The equation of motion for a spherical, steady state wind is
\begin{equation}
\left(v-\frac{c_s^2}{v}\right)\frac{dv}{dr}+\frac{GM_\ast(1-\Gamma_\ast)}{r^2}
+ \frac{2c_s^2}{r}-\frac{dc_s^2}{dr} +F^{rad,l}=0
\end{equation}
where $c_s$ is the gas sound speed. 
We dropped the 'r' subscript for simplicity.
The principles of solving this equation of motion are presented in CAK
(see also PPK). Here we generally follow them to calculate the velocity
law and the stellar mass-loss rate in the zero sound speed case.
Equation (A9) reduces to
\begin{equation}
v\frac{dv}{dr}+\frac{GM_\ast(1-\Gamma_\ast)}{r^2}+F^{rad,l}=0.
\end{equation}
Using the continuity equation and equation (A8), equation (A10) becomes:
\begin{equation}
vv'+\frac{GM_\ast(1-\Gamma_\ast)}{r^2}+\frac{G M \Gamma_\ast}{r^2} f_{FD}(r) 
k \left(\frac{4\pi}{\sigma_e v_{th} \MDOT_\ast}\right)^\alpha (vv')^{\alpha}=0,
\end{equation}
where $v'=\frac{dv}{dr}$.

Assuming $f_{FD}(r)=1$, we have the case considered by CAK. Here we repeat
their well known solutions for the velocity law
\begin{equation}
v(r)=v_{\infty} (1 - r_\ast/r)^\beta.
\end{equation}
where $\beta=0.5$,
\begin{equation}
v_\infty^2=\frac{\alpha}{1-\alpha}\left(\frac{2GM_\ast(1-\Gamma_\ast)}{r_\ast}\right)
\end{equation}
and the mass-loss rate
\begin{equation}
\MDOT_\ast= \frac{4\pi GM_\ast}{\sigma_e v_{th}}
\frac{\alpha}{1-\alpha}(1-\Gamma_\ast)^{-(1-\alpha)/\alpha}
\left( (1-\alpha) k\Gamma_\ast \right)^{1/\alpha}.
\end{equation}

When the star is treated as a point-source,
the optical depth parameter does not vary with radius and is
\begin{equation}
t=\frac{v_{th}\MDOT_\ast\sigma_e}{2\pi v_{\infty}^2r_\ast}
\end{equation} and subsequently
\begin{equation}
M(t)=k \left( \frac{v_{th}\MDOT_\ast\sigma_e}{2\pi v_{\infty}^2r_\ast}\right)^{-\alpha}.
\end{equation}
On the other hand using the critical point conditions (see CAK, PPK and
Gayley 1995) the force multiplier is
\begin{equation}
M(t)=M(t_c)=\frac{1-\Gamma_\ast}{(1-\alpha)\Gamma_\ast},
\end{equation}
where the 'c' subscript indicates that a quantity is evaluated
at the wind critical point.

Comparing the right hand sides of equations (A16) and  (A17) and using 
equation (A13),
$\MDOT_\ast$ can be expressed as
\begin{equation}
\MDOT_\ast=\frac{4\pi GM_\ast(1-\Gamma_\ast)}{\sigma_e v_{th}}\frac{\alpha}{1-\alpha}
       \left(\frac{k}{M(t_c)}\right)^{1/\alpha}.
\end{equation}
Equation (A18) demonstrates the importance of the force multiplier 
at the critical point in determining the mass-loss rate.

For the finite disc case, the mass-loss rate is also given by equation (A18)
but the force multiplier at the critical points is now
\begin{equation}
M(t_c)=\frac{1-\Gamma_\ast}{(1-\alpha)\Gamma_\ast f_{FD,c}}.
\end{equation}
For the velocity law we assume that it can be approximated by equation (A12)
and find that
\begin{equation}
v_\infty^2=\frac{\alpha}{1-\alpha} \frac{GM_\ast(1-\Gamma_\ast)}{r_\ast}
\frac{1}{\beta} (1-r_\ast/r_c)^{-2\beta+1}. 
\end{equation}
We confirm PPK results that this approximation works fairly well for
$\beta=0.8$ and $r_c=1.05r_\ast$.
Note that in this case the force multiplier increases with radius.

Now we move to a problem of solving the equation of motion 
when  $M_{max}< M(t_c)$. In this case we assume that the force
multiplier is constant, i.e., $M(t)=M_{max}$ for all locations in a wind.
Such a simple approach allows the equation of motion to be written as
\begin{equation}
vv'+\frac{GM_\ast(1-\Gamma_\ast)}{r^2}+\frac{G M \Gamma_\ast }{r^2} f_{FD}(r) 
M_{max}=0.
\end{equation}

For $f_{FD}=1$, we find that the velocity law is again as in the  CAK case
with
\begin{equation}
v_\infty^2=
 \frac{2GM_\ast(\Gamma_\ast(1+M_{max})-1)}{r_\ast},
\end{equation}
while the mass-loss rate is
\begin{equation}
\MDOT_\ast=\frac{4\pi GM_\ast(\Gamma_\ast(1+M_{max})-1) }{\sigma_e v_{th}}
       \left(\frac{k}{M_{max}}\right)^{1/\alpha}.
\end{equation}

For the finite disc case, the velocity law is:
\begin{equation}
\begin{array}{ll} 
v(r) & =  \left( \frac{2 G M}{(1+\alpha)r r_\ast^2}
( (1+\alpha)(1-\Gamma_\ast)r_\ast^2
+\Gamma_\ast M_{max} (r^2(1- H2F1(-0.5, -\alpha; 0.5; \frac{r_\ast^2}{r^2}))
\right. \\
 & \left.  
-r_\ast^2 H2F1(0.5, -\alpha; 1.5; \frac{r_\ast^2}{r^2})) 
+ r r_\ast(\Gamma_\ast(1-M_{max}(1-h1-h2))-\alpha (1-\Gamma_\ast)-1)) \right)^{1/2}\\
\end{array}
\end{equation}
where H2F1 is hypergeometric function, 
$h1=H2F1(-0.5, -\alpha; 0.5; 1)$ and 
$h2=H2F1(0.5, -\alpha; 1.5; 1)$.
The terminal velocity is
\begin{equation}
v_\infty=
 \left(\frac{2GM_\ast}{(1+\alpha)r_\ast}(\Gamma_\ast(1-M_{max}(1+\alpha/r_\ast-h1
-h2)) - \alpha(1-\Gamma_\ast)-1)\right)^{1/2}.
\end{equation}
And finally the mass-loss rate is
\begin{equation}
\MDOT_\ast= \frac{4\pi GM_\ast (\Gamma_\ast(1+f_{FD,c}M_{max})-1)}
{\sigma_e v_{th}}
\left(\frac{k}{M_{max}} \right)^{1/\alpha}.
\end{equation}

\end{document}